\documentclass{ws-ijmpcs}

\usepackage[applemac]{inputenc}
\usepackage[T1]{fontenc}
\usepackage{amssymb,amsmath,amsfonts,dsfont}

\def\F{\mathcal{F}}
\def\M{\mathcal{M}}

\def\R{\mathcal{R}}

\def\I{\mathcal{I}}

\def\L{\mathcal{L}}
\def\bk{\mathbf{k}}

\def\TE{\mathrm{TE}}
\def\TM{\mathrm{TM}}

\def\d{\mathrm{d}}

\def\PP{\mathrm{PP}} 
 
\def\PFA{\mathrm{PFA}}
\def\LD{\mathrm{LD}}

\def\txi{\tilde{\xi}}
\def\ixi{\imath \xi}

\def\LOR{\frac{L}{R}}

\def\eps{\varepsilon}
\def\Drud{\mathrm{Drud}}
\def\plas{\mathrm{plas}}

\def\perf{\mathrm{perf}}

\def\max{\mathrm{max}}

\def\lmax{\ell_\max}

\begin{document}

\markboth{Canaguier-Durand, Lambrecht, Reynaud, Gu\'{e}rout}
{The Casimir effect in the sphere-plane geometry}

\catchline{}{}{}{}{}
%
%%%%%%%%%%%%%%%%%%%%%%%%%%%%%%%%%%%%%%%%%%%%%%%%%%%%%%%%%%%%%%%%%%%%

\title{THE CASIMIR EFFECT IN THE SPHERE-PLANE GEOMETRY}

\author{ ANTOINE CANAGUIER-DURAND, ROMAIN GU\'{E}ROUT
%\footnote{ISIS - Groupe des Nanostructures ; 8 allée Gaspard Monge ; BP 70028  ; 67083 Strasbourg Cedex  ; France.}
, PAULO A. MAIA NETO, ASTRID LAMBRECHT, SERGE REYNAUD}

\address{Laboratoire Kastler Brossel, UPMC, CNRS, ENS, \\ 4 place Jussieu ;
 75005 Paris, France\\
antoine.canaguier@gmail.com}

%\author{}
%
%\address{Laboratoire Kastler Brossel, UPMC, CNRS, ENS, \\ 4 place Jussieu ; 
%75005 Paris, France}

\maketitle

\begin{history}
\received{Day Month Year}
\revised{Day Month Year}
\end{history}

\begin{abstract}
We present calculations of the Casimir interaction between a sphere and a plane, using a multipolar expansion of the scattering formula. This configuration enables us to study the nontrivial dependence of the Casimir force on the geometry, and its correlations with the effects of imperfect reflection and temperature. The accuracy of the Proximity Force Approximation (PFA) is assessed, and is shown to be affected by imperfect reflexion. Our analytical and numerical results at ambient temperature show a rich variety of interplays between the effects of curvature, temperature, finite conductivity, and dissipation. 
\end{abstract}

\ccode{PACS numbers: 11.25.Hf, 123.1K}

\section*{Introduction}	

% large view introduction (th-exp comp)
Measuring the Casimir force\cite{casimir1948attraction} has been the aim of an increasing number of experiments 
%\nocite{lamoreaux1997demonstration,mohideen1998precision,harris2000precision,ederth2000template,chan2001quantum,chan2001nonlinear,bressi2002measurement,decca2003measurement,chen2004theory,decca2007tests,munday2007precision,zwol2008influence,munday2009measured,jourdan2009quantitative,masuda2009limits,de2009halving,sushkov2011observation} 
in the past fifteen years (see Ref.~\refcite{lambrecht2011casimir} and references therein).
%\cite{lamoreaux1997demonstration}\cdash\cite{sushkov2011observation} 
The comparison of these measurements with theoretical predictions from Quantum ElectroDynamics have been applied to put constraints on hypothetical new forces predicted by unification models.\cite{onofrio2006casimir,klimchitskaya2009casimir} Accurate theoretical computations, which account for a realistic modeling of experimental conditions, are sorely needed for all comparisons to be reliable.\cite{lambrecht2006casimir,reynaud2010scattering}

% small view introduction (temperature and finite conductivity)
The configuration initially considered by Casimir\cite{casimir1948attraction} is a pair of two infinite parallel plates, perfectly reflecting and at zero temperature, which constitutes a much idealized case. The effect of finite conductivity plays an essential role in the accurate determination of the force,\cite{lambrecht2000casimir} while the thermal fluctuations
%\nocite{bostrom2000thermal,brevik2006thermal,klimchitskaya2006experiment,brevik2008analytical,milton2009recent,ingold2009quantum}
 give rise to a remarkable interplay with the former effect.\cite{bostrom2000thermal}
 %\cite{bostrom2000thermal}\cdash\cite{ingold2009quantum} 
 Indeed, in the calculations performed for the geometry of two parallel plates,
the Casimir force computed from the dissipative Drude model turns out
to be a factor of 2 smaller than the result obtained from the lossless plasma model.

\begin{figure}[htb]
\centerline{\psfig{file=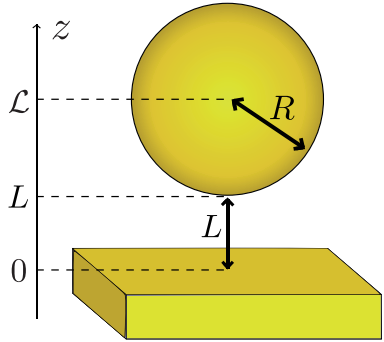,width=7cm}}
%\vspace*{8pt}
\caption{Sphere of radius $R$ and a flat infinite plate at a distance $L$, separated in the $z$-direction. The center-to-plate distance is $\L=L+R$.}
\label{fig:schema_sphereplane}
\end{figure}

% small view introduction (dependance on geometry)
Moreover, the geometry of two parallel infinite plates is quite particular, for example the reflection on the mirrors is specular and the two electromagnetic polarizations are uncoupled. These features are not present in the general case, and the spectral nature of the Casimir effect makes it sensitive to changes in the geometry. 
% presentation of the situation
The nontrivial situation of a sphere and a plane, the configuration of the most precise experiments, is depicted in Figure~\ref{fig:schema_sphereplane}: a sphere of radius $R$ is located at a distance $L$ from an infinite plane along the $z$-direction, $L$ being the distance of closest approach and $\L=L+R$ the center-to-plate distance. In this configuration, not only curvature, but also non-specular reflection, coupling of electromagnetic polarizations and finite size are present. 

\vspace{-0.3cm}

% PFA formulas and beyond-PFA rhoE (+ ratio plasma/Drude 2 at long-distance)
In the sphere-plane configuration, the Casimir force is usually derived from the Casimir energy evaluated in the parallel-plate geometry with an integration over the sphere surface. The resulting proximity force approximation (PFA) for the Casimir force is\cite{deriagin1968kolloid}
\begin{align}\label{F_PFA}
F_\PFA (L,R) = 2\pi R \frac{E_\PP(L)}{A} ~ ,
\end{align} 
where $\frac{E_\PP}{A}$ is the Casimir energy per unit area in the plane-plane configuration. Eq.(\ref{F_PFA}) is expected to provide an accurate description in the limit of small aspect ratios $\LOR$ (see Refs.~\refcite{schaden1998infinity,schaden2000focusing,jaffe2004casimir} for derivations with perfect mirrors at zero temperature). Even if the spheres used in the experiments are much larger than the distance from the plane, mastering the error made by this approximation remains necessary in order to match the experimental accuracy level.\cite{krause2007experimental}
% possible interplay
Moreover, there is no reason why the thermal, finite conductivity, and beyond-PFA corrections should be expected to be independent. Therefore, an accurate description of the sphere-plane configuration has to take these effects into account simultaneously within a single theoretical model.

% Résumé du papier
In this paper, we present a complete analysis of the Casimir interaction in the sphere-plane configuration, using a multipolar expansion of the scattering formula for the evaluation of the force. This enables us not only to assess the accuracy of the PFA quantities, but also to investigate the nontrivial dependence of the Casimir effect on the geometry, and its interplay with the effects of temperature and optical properties of the mirrors. 
% annonce du plan
In the first section, we develop the scattering formula for the sphere-plane configuration and evoke, in the prospect of numerical evaluations, the need for using a finite number of multipoles in the expansion. Section~2 discusses the results at zero temperature, first for perfect and then for metallic mirrors. The ambient temperature case is finally investigated in Section~3. 

\section{Method}

We use the scattering approach\cite{lambrecht2006casimir} to compute the Casimir interaction in the sphere-plane configuration at ambient temperature. The Casimir free-energy is written as a sum over the Matsubara frequencies $\xi_n$, $(n\geq 0)$: 
\begin{align}
\F = k_B T \sum_{n}^{'} \ln\det \left[ \I - \M(\xi_n) \right] ~ ~  , ~ ~ ~ \xi_n = n \frac{2\pi k_B T}{\hbar} \\
\M(\xi) = \R_S(\xi) e^{-\mathcal{K}(\xi)\L} \R_P(\xi) e^{-\mathcal{K}(\xi)\L} ~ ,
\end{align}
where the primed sum means that the $(n=0)$-term is counted for a half. The reflection operators of the sphere,  $\R_S(\xi)$, and the plate, $\R_P(\xi)$, are evaluated with reference points at the sphere center and at its projection on the plane, respectively. The operator  $e^{-\mathcal{K}(\xi) \L}$ accounts for  one-way propagation along the $z$ axis between these  points, separated by the length $\L.$ Thus, the operator $\M (\xi)$  represents one round-trip propagation inside the open cavity formed by the two surfaces. At zero temperature, the discrete set of Matsubara frequencies becomes continuous, and the sum has to be replaced by a continuous integral over all frequencies in $\mathds{R}^+$.

For a wave of imaginary frequency $\omega=\imath \xi$, we first introduce the electromagnetic multipoles basis $\mid \ell, m, P, \rangle$, where $\ell(\ell+1)$ and $m$ denote the usual angular momentum discrete eigenvalues (with $\ell \geq 1$ and $-\ell\leq m \leq \ell$), and $P=E,M$ for electric and magnetic multipoles. This set of modes is well adapted to the spherical symmetry of $\R_S$. The planar wave basis is however better adapted for the planar reflection operator $\R_P$ and the two translations operators $e^{\mathcal{K}(\xi)\L}$, which are diagonal in this representation. We denote those planar modes by the representation $\mid \bk, \phi, p\rangle$, where $\bk$ is the wave-vector component parallel to the $xy$-plane, $\phi=\pm$ is the direction of propagation along the $z$-axis, and $p=\TE,\TM$ the polarization. 

We will thus use both sets of electromagnetic modes, by expressing the round-trip operator $\M(\xi)$ in the multipole basis, inserting the identity in the planar wave basis (see Ref.~\refcite{canaguier2010thermal2} for a more detailed derivation). The expressions of the non-zero diagonal blocks ($m_1=m_2=m$) then read:
\begin{align}
\label{Mintegral}
\M^{(m)}(\xi)_{1,2}  = 
&\int\frac{\d^2\bk}{(2\pi)^2}\sum_{p=\TE,\TM}
\langle \ell_1, m, P_1 | \R_S(\xi) |\bk,+,p \rangle \nonumber \\
&\times r_p(\bk,\xi) e^{-2\kappa \L} \,\langle \bk,-,p \mid \ell_2, m, P_2\rangle 
\end{align}
This expression has a simple interpretation when read from right to left:
a multipole wave $ \mid \ell_2, m,  P_2\rangle$ is first decomposed into plane waves
(coefficients $\langle\bk,-,p~\mid~\ell_2,m,P_2\rangle$) which
propagate towards the plane (factor $e^{-\kappa \L}$). After reflection by the plane
(specular amplitude  $r_p(\bk,\xi)$), the plane wave components
 propagate back to the sphere (second factor  $e^{-\kappa \L}$) and are finally scattered
into a new multipole wave $\mid \ell_1, m, P_1\rangle.$

% conclu
Eq.(\ref{Mintegral}) gives an exact formula for the expression of the Casimir free-energy and its various derivatives in the sphere-plane geometry, valid for arbitrary values of all the parameters. However, in order to evaluate numerically this quantity, one has to truncate the dimension of the operator $\M$. This can be done by setting a maximum value for the quantum number $\ell \leq \lmax$. 
% numerical truncation
From the localization principle, the value of $\lmax$ required to obtain a numerical evaluation for a given accuracy level is expected to scale with the dimensionless parameter $\txi=\xi \frac{R}{c}$. As the frequencies giving the main contribution to the Casimir effect scale as $\xi \lesssim \frac{c}{L}$, it follows that the required $\lmax$ should scale as $\frac{R}{L}$.

%\vspace{-0.3cm}

\begin{figure}[htb]
\centerline{\psfig{file=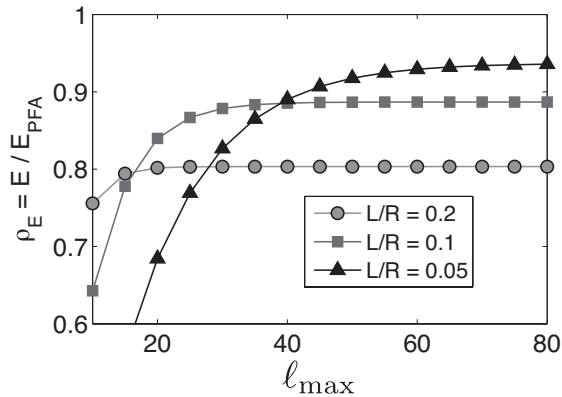,width=8cm}}
%\vspace*{8pt}
\caption{Casimir energy $E$ computed with the scattering formula, normalized by the PFA quantity $E_\PFA$, as a function of the cut-off $\lmax$, for different values of the aspect ratio $\LOR$.}
\label{fig:rhoE_T0_vs_lmax}
\end{figure}

%\vspace{-0.5cm}

 Figure \ref{fig:rhoE_T0_vs_lmax} confirms this effect of the truncation by presenting the Casimir energy at zero temperature $E$, normalized by the PFA result $E_\PFA$. We first consider a sphere that is five times larger than the distance to the plane ($\LOR=0.2$, circles). When the truncation value $\lmax$ increases, the result of the numerical evaluation quickly converges for the value $\lmax=20$, approximatively. Then, for a twice bigger sphere ($\LOR=0.1$, squares), the same convergence is observed, but it is twice slower, such that accurate evaluations need $\lmax \gtrsim 40$. This proportional behavior is observed as well for even larger spheres ($\LOR=0.05$, triangles).

From this observation we can conclude that for a given $\lmax$ in the numerical evaluation, results are only accurate for aspect ratios $\LOR$ larger than a minimum value, which is inversely proportional to $\lmax$. This multipolar treatment is then well-suited for situations where the aspect ratio takes intermediate and large values $(\LOR \gtrsim 1)$. However, the use of great values for $\lmax$ in the numerics can also enable evaluations in the opposite regime, where $\LOR$ takes smaller values.

\section{Results at zero temperature}

\subsection{Perfect mirrors}

% comparison with PFA at short L/R
We start our analysis with the simplest case of perfectly reflecting mirrors at zero temperature. Only two length scales are involved in this case, the sphere radius $R$ and the distance $L$ between the two objects. The dependance of the Casimir effect is then strictly geometrical, and the important parameter is the aspect ratio $\frac{L}{R}$. 

We compare the numerical results for the Casimir energy $E$ obtained from the scattering formalism to the energy computed within PFA, $E_\PFA$. The correction factor $\rho_E=\frac{E}{E_\PFA}$ is then a function of $\LOR$ only, and does not depend separately on $L$ and $R$. For a vanishingly small value of $\LOR$, the domain of validity for the PFA should be recovered and the ratio $\rho_E$ should then go to unity. Assuming a Taylor expansion of $\rho_E$ for small values of $\LOR$:
\begin{align}\label{def_betaE}
\rho_E = \frac{E}{E_\PFA} = 1 + \beta_E \LOR + \gamma_E \left( \LOR \right)^2 + \cdots ~ ,
\end{align}
a usual way to assess the accuracy of the approximated quantities $E_\PFA$ is to estimate the linear correction coefficient $\beta_E$. A similar treatment for the Casimir force $F=-\frac{\partial E}{\partial L}$ or force gradient $G=-\frac{\partial F}{\partial L}$ leads to the linear coefficients $\beta_F$ and $\beta_G$. Scalar derivations\cite{schaden1998infinity,bulgac2006scalar,bordag2008casimir} give the estimation $\beta_E = \frac{1}{3} - \frac{5}{\pi^2} \simeq -0.173$. An experimental study\cite{krause2007experimental} has also studied this linear correction term and has put a constraint for the force gradient $\left| \beta_G \right| \leq 0.4$.

% figure rho_E_T0_perfect_fit
%\vspace{-0.5cm}

\begin{figure}[htb]
\centerline{\psfig{file=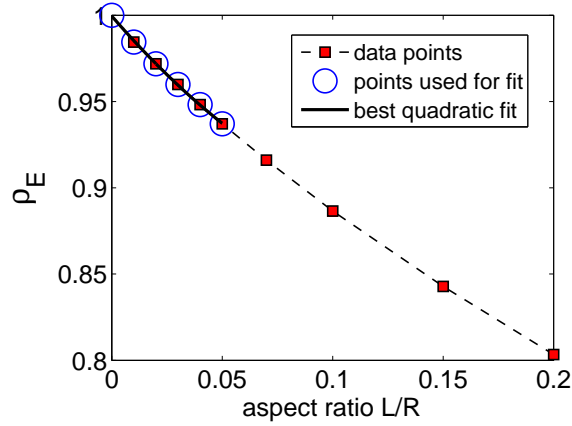,width=8cm}}
%\vspace*{8pt}
\caption{Correction factor $\rho_E=\frac{E}{E_\PFA}$ for the Casimir energy, as a function of the aspect ratio $\LOR$. The mirrors are perfectly reflecting at zero temperature. Red squares represent the numerically obtained data with the scattering method, the blue circles indicate the points used to compute the extrapolating polynomial, presented by a solid line.}
\label{fig:rhoE_T0_perfect_fit}
\end{figure}

% présenter la méthode de fit
In Fig.~\ref{fig:rhoE_T0_perfect_fit} we present the numerically obtained values for this ratio with red squares. They lie below unity, meaning that PFA always overestimates the magnitude of the energy, and indeed tend to unity as the aspect ratio vanishes. In order to estimate the linear coefficient $\beta_E$ from this curve, we use the leftmost points (blue circles) to extrapolate the data to unity, with the help of a quadratic fit (solid curve). We finally extract the initial slope from this fit, which is the linear correction coefficient $\beta_E$.

% donner résultats numériques (polynome, beta)
With this procedure, we obtain $\beta_E \simeq - 1.47$, a number eight times larger than the scalar result, however in good agreement with the results of Refs.~\refcite{emig2008fluctuation,maia2008casimir}. As the scattering formalism considers the electromagnetic field with the two polarizations simultaneously, we conclude that the coupling of polarizations plays an important role in the Casimir effect between a sphere and a plane. The corresponding linear coefficient for the Casimir force gradient $\beta_G = \frac{\beta_E}{3} \simeq -0.49$ lies outside the experimental bound of $0.4$ and would seem to be in contradiction with it. However, finite conductivity effects play a significant role in the experiment, so we need to consider the full theoretical model for real metals, discussed in the next subsection, in order to compare with the experimental bound.

\subsection{Metallic mirrors}

The next step in the realistic description of the experimental configuration is the inclusion of the effects of finite conductivity on the mirrors. This can be done through the plasma and Drude dielectric functions, 
\begin{align}\label{eps_drude}
&\eps^\plas(\ixi) = 1 + \frac{\omega_P^2}{\xi^2} 
&\eps^\Drud(\ixi) = 1 + \frac{\omega_P^2}{\xi (\xi + \gamma)}
\end{align}
expressed at imaginary frequencies $\omega=\ixi$, where $\omega_P$ is the plasma frequency and $\gamma$ is the relaxation frequency associated to dissipation. The plasma dielectric function $\eps^\plas$ is obtained from Eq.~\ref{eps_drude} in the lossless limit $\gamma \rightarrow 0$. 

\begin{figure}[htb]
\centerline{\psfig{file=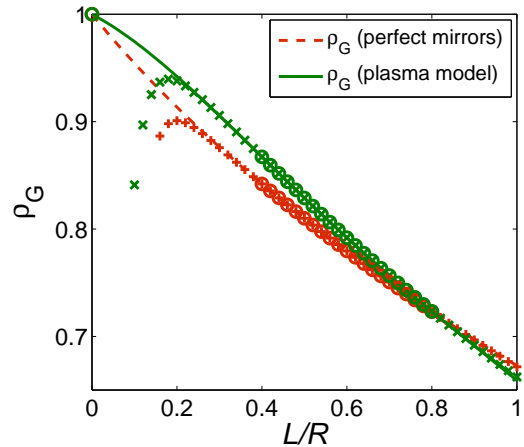,width=7.5cm}}
%\vspace*{8pt}
\caption{Correction factor $\rho_G=\frac{G}{G_\PFA}$ for the Casimir force gradient at zero temperature, as a function of the aspect ratio $\LOR$. The perfectly reflecting case is presented in red, the case of mirrors described by the plasma model in green. Crosses represent the numerical data, accurate for $\LOR \gtrsim 0.4$. The data points used for the fit are specified with circles, and the red dashed-line and green solid line are the obtained polynomial fits. The plasma wavelength is $\lambda_P = 136~$nm.}
\label{fig:rhoG_T0_bending}
\end{figure}

% nanosphere of 100nm to see effects of imperfect reflection
We then conduct the same procedure as in the previous case but with mirrors described by the plasma model\cite{canaguier2009casimir}. We study the Casimir force gradient $G$ in the case of a nanosphere with radius $R=100$~nm, in order to be able to observe the effects of imperfect reflection at short distances $(L \lesssim \lambda_P=\frac{2\pi c}{\omega_P})$. Fig.~\ref{fig:rhoG_T0_bending} presents the obtained results, for perfectly reflecting mirrors in red and for the plasma model in green.  

%\vspace{-0.5cm}

% présenter figure bending

% curve changes -> accuracy affected
First, the fact that the two curves do not superimpose means that the accuracy of PFA is affected by the introduction of the finite conductivity in the mirrors. The two curves do not have the same shape at small distances: while in the case of perfect mirrors $\rho_G$ is always a convex function of $\LOR$, the curve corresponding the plasma model (green curve in Fig.~\ref{fig:rhoG_T0_bending}) shows an inflection point and becomes concave for small values of the aspect ratio. This bending lowers the absolute value of the linear correction coefficient, which becomes $\beta_G \simeq -0.2$ when using the plasma model for the description of the mirrors. This number is back in agreement with the experimental bound $\left| \beta_G \right| <0.4$, and is a sign of correlations between the effects of geometry and finite conductivity. 
 
% bending -> changes beta_G

% -0.2 -> no more in disagreement with beta exp.

% correlations

\section{Results at ambient temperature}

\subsection{Perfect mirrors}

We now include the thermal fluctuations to the vacuum fluctuations of the electromagnetic field, through the summation over Matsubara frequencies. In order to observe the effect of temperature on the Casimir force, we introduce the thermal correction factor\cite{canaguier2010thermal} for the Casimir force $\vartheta=\frac{F(T)}{F(0)}$, and first consider the case of perfect mirrors. A similar factor $\vartheta_\PFA$ can be derived for PFA quantities.%, and is presented with a dashed-line on Fig.~\ref{fig:vartheta_F_perfect}.

\begin{figure}[htb]
\centerline{\psfig{file=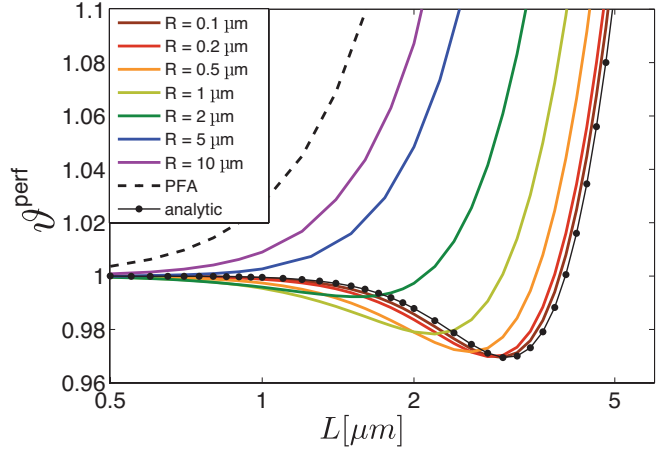,width=9.5cm}}
%\vspace*{8pt}
\caption{Thermal correction factor $\vartheta=\frac{F(T)}{F(0)}$ for the Casimir force in the sphere-plane geometry, as a function of the distance $L$. The PFA result $\vartheta_\PFA$ is recalled by a dashed-line, the solid lines are the results computed from the scattering formula, with different colors for different sphere radii. The analytical limit of small spheres $(R\ll L)$ is drawn as a dotted-line. The temperature is $T=300~$K.}
\label{fig:vartheta_F_perfect}
\end{figure}

The thermal correction factors $\vartheta$ obtained for different values of the sphere radius $R$ are presented in Fig.~\ref{fig:vartheta_F_perfect} with different colors. The first observation is that we always have $\vartheta \leq \vartheta_\PFA$, meaning that the PFA quantities overestimate the contribution of thermal photons to the Casimir force. Moreover, a correct treatment of the geometry enables us to observe a non-trivial $R$-dependence for the thermal correction factor $\vartheta_F$. This is is a clear sign of correlations between the effects of temperature and geometry, which have been observed for scalar fields.\cite{klingmuller2008geothermal}

For small spheres ($R \lesssim 2~\mu$m), we observe that $\vartheta <1$ over a large range of distances, a feature that never appears for the PFA quantities. In such case the thermal photons have a repulsive contribution to the Casimir force, which can be associated with the appearance of negative values for the Casimir entropy\cite{canaguier2010thermal2} $S=-\frac{\partial \F}{\partial T}$, where $\F$ is the Casimir free-energy. 

This observation is confirmed by the analytical result for the small sphere limit ($R \ll L$) that can be obtained from the corresponding long-distance limit (LD) of the Casimir free-energy, introducing the dimensionless parameter $\nu = \frac{2\pi \L}{\lambda_T}$ :
\begin{align}\label{vartheta_F_analytical}
\F^\perf_\LD = - \frac{3 \hbar c R^3}{4 \lambda_T \L^3} \phi(\nu) ~ ~ ~ \mbox{with} ~  ~ ~ 
\phi(\nu) = \frac{\nu \sinh \nu + \cosh \nu \left( \nu^2 + \sinh^2 \nu \right)}{2 \sinh^3 \nu} ~ ~ ,
\end{align}
presented with a dotted-line on Fig.~\ref{fig:vartheta_F_perfect}. Let us note that the low-temperature limit $(R \ll L \ll \lambda_T)$ of Eq.(\ref{vartheta_F_analytical}) agrees with the corresponding limit in Ref.~\refcite{bordag2010vacuum}.

\subsection{Metallic mirrors}

Finally, we include altogether in the model realistic descriptions for the geometry, temperature and optical response for metallic mirrors, by computing the Casimir force $F$ for metallic mirrors at ambient temperature. We concentrate on the effect of dissipation on the Casimir force, by studying the ratio $\frac{F^\plas}{F^\Drud}$ of the force for dissipative and non-dissipative metals. In the case of two parallel plates at non-zero temperature, as well as for the derived PFA quantities, this ratio shifts from unity at small distances to the value $2$ at large distances.

\begin{figure}[htb]
\centerline{\psfig{file=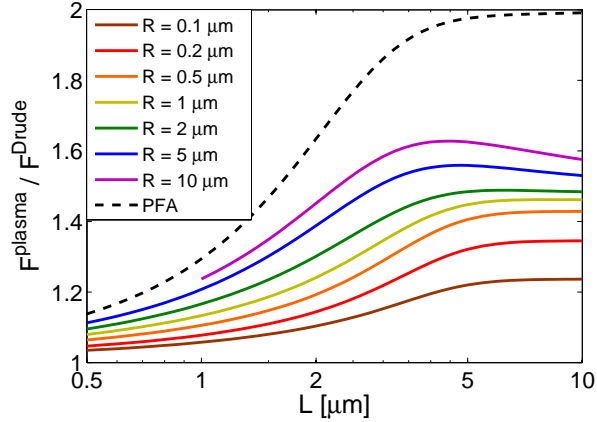,width=8.5cm}}
%\vspace*{8pt}
\caption{Ratio of the Casimir force $\frac{F^\plas}{F^\Drud}$ for lossless plasma and dissipative Drude models with respect to the distance $L$. The PFA result is recalled with a dashed-line, the solid lines are the results computed from the scattering formula, with different colors for different sphere radii. The temperature is $T=300~$K, the plasma wavelength is $\lambda_P=136~$nm and the length scale associated to the dissipation is $\lambda_\gamma = \frac{2\pi c}{\gamma} = 34~\mu$m.}
\label{fig:vartheta_F_perfect}
\end{figure}

In Fig.~\ref{fig:vartheta_F_perfect} we present this ratio as a function of the distance $L$ for different sizes of the sphere. The curves differ from the one obtained from PFA quantities, which is a sign of correlations between the effects of geometry, temperature and dissipation. Moreover, the long-distance limit of this ratio is no more $2$ as in the plate-plate configuration, but a rather an increasing function of the dimensionless parameter $\alpha=2\pi \frac{R}{\lambda_P}$, that can be derived analytically:\cite{canaguier2010thermal2}
\begin{align}
f(\alpha) = \frac{3}{2} \left( 1+ \frac{1}{\alpha^2} - \frac{\coth \alpha}{\alpha} \right) ~ ,
\end{align}
taking values between $1$ and $\frac{3}{2}$. This observation hints at the existence of a strong interplay between the geometric, thermal, and dissipative corrections. It also means that the gap between the results obtained from plasma and Drude model is much smaller than expected from the PFA quantities.

\section*{Conclusions}

In this paper we have investigated the Casimir interaction in the sphere-plane configuration, which can be seen as a simple example for studying the dependence of the Casimir effect on the geometry. The scattering approach allows to treat exactly this configuration, taking into account both the non-zero temperature and the metallic nature of reflectors.

First, we have observed that the coupling of polarizations plays a role in the Casimir effect, a complete electromagnetic treatment is thus necessary to get reliable evaluations in this geometry. The inclusion of a realistic description for the material properties also affects the accuracy of the quantities obtained from the PFA, and it results in a much smaller linear correction coefficient $\beta_G$ which lies within the experimental bound. 

At ambient temperature, the correlations that we observe between the effect of geometry and temperature can lead to a repulsive contribution of thermal photons to the Casimir force, and to negative values for the entropy. Finally we discuss the effect of dissipation on the materials, and the long-distance ratio between the force obtained from plasma model and Drude model, which is $2$ in the parallel-plate geometry, is reduced to, at most, $3/2$. 

\section*{Acknowledgments}

The authors thank  I. Cavero-Pelaez and  G.-L. Ingold  for
discussions, CAPES-COFECUB and the French Contract
ANR-06-Nano-062 for financial support, and the ESF Research Networking Programme CASIMIR
(www.casimir-network.com) for providing excellent opportunities for
discussions on the Casimir effect and related topics. P.A.M.N. thanks H.M. Nussenzveig for discussions and  CNPq and Faperj for financial support.

\bibliography{biblio}
\bibliographystyle{phjcp}

%%\begin{thebibliography}{000} %for 3 digits
%%\begin{thebibliography}{00}  %for 2 digits
%\begin{thebibliography}{0}    %for 1 digit
%
%%%journal paper
%\bibitem{jpap} R. Loren and D. B. Benson, {\it J. Comput. 
%System Sci.} {\bf 27}, 400 (1983).
%
%%%collaboration
%\bibitem{colla} OPAL Collab. (G. Abbiendi {\it et al}.), 
%{\it Eur. J. Phys. C\/} {\bf 11}, 217 (1999).
%
%%%normal book (authors)
%\bibitem{autbk} R. Loren and D. B. Benson, {\it Introduction to String 
%Field Theory}, 2nd edn. (Springer-Verlag, New York, 1999).
%
%%%normal book (editors)
%\bibitem{edbk} R. Loren and D. B. Benson (eds.), {\it Introduction to 
%String Field Theory}, 2nd edn. (Springer-Verlag, New York, 1999).
%
%%%review volume
%\bibitem{rvo} C. M. Wang, J. N. Reddy and K. H. Lee, New set of
%buckling parameters, in {\it Shear Deformable Beams}, ed.~T. Rex 
%(Elsevier, Oxford, 2000), p.~201.
%
%%%book in a series
%\bibitem{seri} R. Loren, J. Li and D. B. Benson, Deterministic flow-chart 
%interpretations, in {\it Introduction to String Field Theory},  
%Ad. Series in Math. Phys., Vol.~3 (Springer-Verlag, New York, 1999), 
%p.~401.
%
%%%proceedings
%\bibitem{pro} R. Loren, J. Li and D. B. Benson, Deterministic
%flow-chart interpretations, in {\it Proc. 3rd Int. Conf. 
%Entity-Relationship Approach}, eds. C. G. Davis and R. T. Yeh 
%(North-Holland, Amsterdam, 1983), p.~421.
%
%%%to be published
%\bibitem{publ} R. Loren, J. Li and D. B. Benson, Deterministic
%flow-chart interpretations, to appear in {\it J. Comput. System Sci.} 
%\end{thebibliography}

\end{document}